\documentclass[apj]{emulateapj}

\newcommand{\kms}{\ifmmode{~{\rm km~s^{-1}}}\else{~km s$^{-1}$}\fi}
\newcommand{\cubecm}{\ifmmode{~{\rm cm^{-3}}}\else{~cm$^{-3}$}\fi}
\newcommand{\lsim}{\lower0.3em\hbox{$\,\buildrel <\over\sim\,$}}
\newcommand{\gsim}{\lower0.3em\hbox{$\,\buildrel >\over\sim\,$}}

\newcommand{\sfr}{M$_\odot$ yr$^{-1}$ Mpc$^{-3}$}

\newcommand{\enzo}{{\sl Enzo}}
\newcommand{\Ms}{\ifmmode{M_\odot}\else{$M_\odot$}\fi}

\newcommand{\hh}{H$_2$}

\newcommand{\tcool}{$t_{\rm{cool}}$}

\newcommand{\tdyn}{$t_{\rm{dyn}}$}

\newcommand{\tvir}{\ifmmode{T_{\rm{vir}}}\else{$T_{\rm{vir}}$}\fi}

\newcommand{\lya}{Ly$\alpha$}
\newcommand{\jj}{\ifmmode{J_{21}}\else{$J_{21}$}\fi}
\newcommand{\flw}{\ifmmode{F_{LW}}\else{$F_{LW}$}\fi}
\newcommand{\tv}{$\langle T \rangle_{\rm v}$}
\newcommand{\tm}{$\langle T \rangle_{\rm m}$}

\begin{document}

\shorttitle{THE START OF REIONIZATION}
\shortauthors{WISE \& ABEL}

\title{How Very Massive Metal Free Stars Start Cosmological Reionization}
\author{John H. Wise\altaffilmark{1,2} and Tom Abel\altaffilmark{1}}

\altaffiltext{1}{Kavli Institute for Particle Astrophysics and
  Cosmology, Stanford University, Menlo Park, CA 94025}
\altaffiltext{2}{Laboratory for Observational Cosmology, NASA Goddard
  Space Flight Center, Greenbelt, MD 21114}
\email{john.h.wise@nasa.gov}

\begin{abstract}

  The initial conditions and relevant physics for the formation of the
  earliest galaxies are well specified in the concordance cosmology.
  Using ab initio cosmological Eulerian adaptive mesh refinement
  radiation hydrodynamical calculations, we discuss how very massive
  stars start the process of cosmological reionization. The models
  include non-equilibrium primordial gas chemistry and cooling
  processes and accurate radiation transport in the Case B
  approximation using adaptively ray traced photon packages, retaining
  the time derivative in the transport equation. Supernova feedback is
  modeled by thermal explosions triggered at parsec scales. All
  calculations resolve the local Jeans length by at least 16 grid
  cells at all times and as such cover a spatial dynamic range of
  $\sim$10$^6$. These first sources of reionization are highly
  intermittent and anisotropic and first photoionize the small scales
  voids surrounding the halos they form in, rather than the dense
  filaments they are embedded in. As the merging objects form larger,
  dwarf sized galaxies, the escape fraction of UV radiation decreases
  and the \ion{H}{2} regions only break out on some sides of the
  galaxies making them even more anisotropic.  In three cases, SN
  blast waves induce star formation in overdense regions that were
  formed earlier from ionization front instabilities.  These stars
  form tens of parsecs away from the center of their parent DM halo.
  Approximately 5 ionizing photons are needed per sustained ionization
  when star formation in 10$^6$ \Ms~halos are dominant in the
  calculation.  As the halos become larger than $\sim$$10^7 \Ms$, the
  ionizing photon escape fraction decreases, which in turn increases
  the number of photons per ionization to 15--50, in calculations with
  stellar feedback only.  Radiative feedback decreases clumping
  factors by 25 per cent when compared to simulations without star
  formation and increases the average temperature of ionized gas to
  values between 3,000 and 10,000 K.
  
\end{abstract}


\section{MOTIVATION}

It is clear that quasars are not responsible to keep the universe
ionized at redshift 6. The very brightest galaxies at those redshifts
alone also provide few photons.  The dominant sources of reionization
so far are observationally unknown despite remarkable advances in
finding sources at high redshift \citep[e.g.][]{Shapiro86, Bouwens04,
  Fan06, Thompson07, Eyles06} and hints for a large number of
unresolved sources at very high redshifts \citep{Spergel07,
  Kashlinsky07} which is still a topic of debate \citep{Cooray07,
  Thompson07}. At the same time, ab initio numerical simulations of
structure formation in the concordance model of structure formation
have found that the first luminous objects in the universe are formed
inside of cold dark matter (CDM) dominated halos of total masses $2
\times 10^5 - 10^6 \Ms$ \citep{Haiman96, Tegmark97, Abel98}.  Fully
cosmological ab initio calculations of \citet{Abel00, Abel02} and more
recently \citet{Yoshida06} clearly show that these objects will form
isolated very massive stars. Such stars will be copious emitters of
ultraviolet (UV) radiation and are as such prime suspect to get the
process of cosmological reionization started.  In fact, one
dimensional calculations of \citet{Whalen04} and \citet{Kitayama04}
have already argued that the earliest \ion{H}{2} regions will
evaporate the gas from the host halos and that in fact most of the UV
radiation of such stars would escape into the intergalactic
medium. Recently, \citet{Yoshida07a} and \citet{Abel07} demonstrated
with full three-dimensional radiation hydrodynamical simulations that
indeed the first \ion{H}{2} regions break out of their host halos
quickly and fully disrupt the gaseous component of the cosmological
parent halo.  All of this gas finds itself radially moving away from
the star at $\sim30\kms$ at a distance of $\sim100$ pc at the end of
the stars life. At this time, the photo-ionized regions have now high
electron fractions and little destructive Lyman-Werner band radiation
fields creating ideal conditions for molecular hydrogen formation
which may in fact stimulate further star formation above levels that
would have occurred without the pre-ionization. Such conclusion have
been obtained in calculations with approximations to multi dimensional
radiative transfer or one dimensional numerical models
\citep{Ricotti02a, Nagakura05, OShea05, Yoshida06, Ahn07,
  Johnson07}. These early stars may also explode in supernovae and
rapidly enrich the surrounding material with heavy elements, deposit
kinetic energy and entropy to the gas out of which subsequent
structure is to form. This illustrates some of the complex interplay
of star formation, primordial gas chemistry, radiative and supernova
feedback and readily explains why any reliable results will only be
obtained using full ab initio three dimensional hydrodynamical
simulations. In this paper, we present the most detailed such
calculations yet carried out to date and discuss issues important to
the understanding of the process of cosmological reionization.

It is timely to develop direct numerical models of early structure
formation and cosmological reionization as considerable efforts are
underway to
\begin{enumerate}
\item Observationally find the earliest galaxies with the James Webb
  Space Telescope \citep[JWST;][]{Gardner06} and the Atacama Large
  Millimeter Array \citep[ALMA;][]{Wilson05},
\item Further constrain the amount and spatial non-uniformity of the
  polarization of the cosmic microwave background radiation
  \citep{Page07},
\item Measure the surface of reionization with LOFAR
  \citep{Rottgering06}, MWA \citep{Bowman07}, GMRT \citep{Swarup91}
  and the Square Kilometer Array \citep[SKA;][]{Schilizzi04}, and
\item Find high redshift gamma ray bursts with SWIFT \citep{Gehrels04}
  and their infrared follow up observations.
\end{enumerate}

We begin by describing the cosmological simulations that include
primordial star formation and accurate radiative transfer.  In
\S\ref{sec:SF}, we report the details of the star formation
environments and host halos in our calculations.  Then in
\S\ref{sec:reion}, we describe the resulting start of cosmological
reionization, and investigate the environments in which these
primordial stars form and the evolution of the clumping factor.  We
compare our results to previous calculations and further describe the
nature of the primordial star formation and feedback in
\S\ref{sec:discussion}.  Finally we summarize our results in the last
section.


%
%

\begin{deluxetable*}{lccccccc}
\tablecolumns{8}
\tabletypesize{}
\tablewidth{\textwidth}
\tablecaption{Simulation Parameters\label{tab:sims}}

\tablehead{
  \colhead{Name} & \colhead{$l$} & \colhead{Cooling model} &
  \colhead{SF} & \colhead{SNe} & \colhead{N$_{\rm{part}}$} &
  \colhead{N$_{\rm{grid}}$} & \colhead{N$_{\rm{cell}}$} \\
  \colhead{} & \colhead{[Mpc]} & \colhead{} & \colhead{} & \colhead{}
  & \colhead{} & \colhead{} & \colhead{} 
} 
\startdata

SimA-Adb & 1.0 & Adiabatic & No & No & 2.22 $\times$ 10$^7$ & 30230 & 9.31
$\times$ 10$^7$ (453$^3$) \\
SimA-HHe & 1.0 & H, He & No & No & 2.22 $\times$ 10$^7$ & 40601 & 1.20
$\times$ 10$^8$ (494$^3$) \\
SimA-RT & 1.0 & H, He, \hh & Yes & No & 2.22 $\times$ 10$^7$ & 44664 & 1.19
$\times$ 10$^8$ (493$^3$) \\
SimB-Adb & 1.5 & Adiabatic & No & No & 1.26 $\times$ 10$^7$ & 23227 & 6.47
$\times$ 10$^7$ (402$^3$) \\
SimB-HHe & 1.5 & H, He & No & No & 1.26 $\times$ 10$^7$ & 21409 & 6.51
$\times$ 10$^7$ (402$^3$) \\
SimB-RT & 1.5 & H, He, \hh & Yes & No & 1.26 $\times$ 10$^7$ & 24013 & 6.54
$\times$ 10$^7$ (403$^3$) \\
SimB-SN & 1.5 & H, He, \hh & Yes & Yes & 1.26 $\times$ 10$^7$ & 24996 & 6.39
$\times$ 10$^7$ (400$^3$)

\enddata
\tablecomments{Col. (1): Simulation name. Col. (2): Box
  size. Col. (3): Cooling model.  Col. (4): Star formation. Col. (5):
  Supernova feedback. Col. (6): Number of dark matter
  particles. Col. (7): Number of AMR grids. Col. (8): Number of unique
  grid cells.}
\end{deluxetable*}

\section{RADIATION HYDRODYNAMICAL SIMULATIONS}
\label{sec:simulations}

We use radiation hydrodynamical simulations with a modified version of
the cosmological AMR code \enzo~to study the radiative effects from
the first stars \citep{Bryan97, Bryan99}.  We have integrated adaptive
ray tracing \citep{Abel02b} into the chemistry, energy, and
hydrodynamics solvers in \enzo~that accurately follow the evolution of
the \ion{H}{2} regions from stellar sources and their relevance during
structure formation and cosmic reionization.

Seven different simulations are discussed here.  Table~\ref{tab:sims}
gives an overview of the parameters and the physics included in these
calculations.  We perform two cosmological realizations, Sim A and B,
with three sets of assumptions about the primordial gas chemistry.
The simplest calculations here assume only adiabatic gas physics and
provide the benchmark against which the more involved calculations are
compared.  We compare this to one model with atomic hydrogen and
helium cooling only and one that includes \hh~cooling.  Massive,
metal-free star formation is included only in the \hh~cooling models.

These calculations are initialized at redshift\footnote{To simplify
  the discussion, simulation A will always be quoted first with the
  value from simulation B in parentheses.} $z$ = 130 (120) when the
intergalactic medium has a temperature of 325 (280) K in box sizes 1
comoving Mpc (1.5 Mpc) for Sim A (B) with three initial nested AMR
grids.  We use the COSMICS package to create the initial conditions,
which uses a Zel'dovich approximation \citep{Bertschinger95,
  Bertschinger01}.  We use the cosmological parameters of
$(\Omega_B\,h^2, \ \Omega_M, \ h, \ \sigma_8, \ n)=(0.024,\ 0.27,\
0.72,\ 0.9,\ 1)$ from first year WMAP results, where the constants
have the usual meaning \citep{Spergel03}.  The changes in the third
year WMAP results \citep{Spergel07} does not affect the evolution of
individual halos studied here but only delays structure formation by
$\sim$40\% \citep{Alvarez06b}.  The adiabatic simulations as well as
the atomic hydrogen and helium cooling only calculations are described
in \citet{Wise07a}. The new models presented here have the exact same
setup and random phases in the initial density perturbation and only
differ in that they include star formation as well as follow the full
radiation hydrodynamical evolution of the \ion{H}{2} regions and
supernova feedback in Sim B.  We use the designations RT and SN to
distinguish cases in which only star formation and radiation transport
were included (RT) and the one model which also includes supernovae
(SN) in Sim B.  We use the same refinement criteria as in our previous
work, where we refine if the DM (gas) density becomes three times
greater than the mean DM (gas) density times a factor of $2^l$, where
$l$ is the AMR refinement level.  We also refine to resolve the local
Jeans length by at least 16 cells.  Cells are refined to a maximum AMR
level of 12 that translates to a spatial resolution of 1.9 (2.9)
comoving parsecs.  This spatial resolution of $\sim$0.1 proper pc is
required to model the formation of the D-type front at small scales
correctly.  Refinement is restricted to the innermost initial nested
grid that has a side length of 250 (300) comoving kpc.

\begin{figure}[b]
\begin{center}
\epsscale{1.15}
\plotone{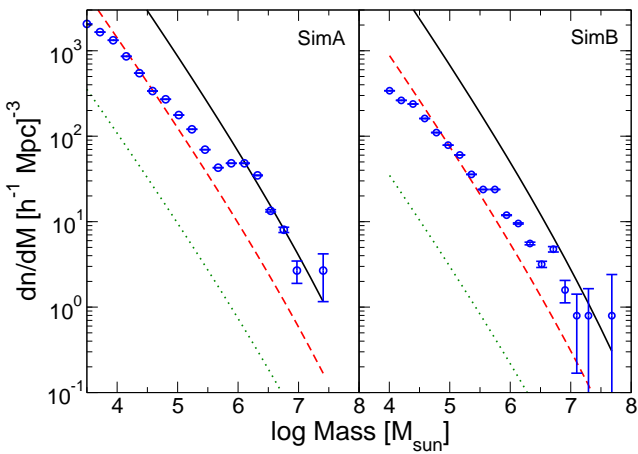}
\caption{\label{fig:massfn} Halo mass functions for Simulation A at $z
  = 15.9$ (left) and B at $z = 16.8$ (right).  The open circles are
  computed from the simulations.  The solid line is the halo mass
  function from the ellipsoidal variant of Press-Schechter formalism
  \citep{Sheth02}.  The dotted line is corrected by a factor of ($L$/1
  Mpc)$^3$, where $L$ = 250 (300) kpc is the comoving length of the
  inner refined region.  The dashed line is corrected by a factor of
  $b$($L$/1 Mpc)$^3$, where $b$ = 35 (20) is the overdensity of halos
  in the biased, refined region where we allow stars to form.  Our
  simulations agree well with the uncorrected S-T function (solid)
  above $\sim10^{6.5}\Ms$ and the corrected function (dashed) below
  this mass scale.}
\end{center}
\end{figure}

These simulations focus on a highly biased region at $z > 15$, which
could be subject to numerical artifacts created by incorrect second-
and higher-order growing modes (i.e. transients) associated with the
Zel'dovich approximation \citep{Scoccimarro98, Crocce06}.  This
suppresses the tails of the density and velocity probability
distributions that leads to rare events (i.e. halos) becoming less
common than expected and is usually avoided by starting at very high
redshifts.  Although this may be the case, the halos examined in our
simulations are well within cosmic variance.  To illustrate this, we
plot the DM halo mass function of the entire simulation volume and
compare it to the one computed by an ellipsoidal variant (S-T) of
Press-Schechter formalism \citep{Press74, Sheth02} in Figure
\ref{fig:massfn}.  The circles represent the data and its associated
Poisson errors, and the lines are the S-T number densities.  The halo
mass function matches well with the S-T function (solid line) above
$\sim10^{6.5} \Ms$.  Below this mass, it decreases relative to the PS
function because our high-resolution region samples a central box with
a side $L$ = 250 (300) comoving kpc.  To correct for this effect, we
multiply the S-T mass function by ($L$/1 Mpc)$^3$ to obtain the dotted
line.  We chose this region because it is highly biased and should
expect the halo mass function to be greater than the cosmic average.
The halo mass function below $\sim10^{6.5} \Ms$ matches well with S-T
function corrected by a factor of $b(L/1\; \rm{Mpc})^3$, where $b$ =
35 (20) is the overdensity of halos in the refined region, that is
plotted as the dotted line in Figure \ref{fig:massfn}.  We note that
recent high-resolution simulations have found that the S-T halo mass
function overestimates rare objects by up to 50\% at all times and
halo abundances at very high redshifts \citep{Iliev06, Reed07,
  Lukic07}.  Transients from the Zel'dovich approximation used in our
initial setup should not affect our general conclusions on the star
formation rate and resulting reionization; however we caution that the
simulations presented here probably underestimates the clumping factor
in the intergalactic medium (IGM) and number densities of low-mass
halos.  The effects of transients in simulations that focus highly
biased regions should be examined more carefully in the future.

%
%

 
The star formation recipe and radiation transport are detailed in
\citet{Wise07c}.  Here we overview the basics about our method.  Star
formation is modelled using the \citet{Cen92b} algorithm with the
additional requirement that an \hh~fraction of $5 \times 10^{-4}$ must
exist before a star forms.  We allow star formation to occur in the
Lagrangian volume of the surrounding region out to three virial radii
from the most massive halo at $z = 10$ in the dark matter only runs as
discussed in \citet{Wise07a}.  This volume that has a side length of
195 (225) comoving kpc at $z = 30$ and 145 (160) comoving kpc at the
end of the calculation.  The calculations with SNe use a stellar mass
$M_\star$ of 170\Ms, whereas the ones without SNe use a mass of
100\Ms.  The ionizing luminosities are taken from no mass loss models
of \citet{Schaerer02}, and we employ the SN energies from
\citet{Heger02}.  Star particles after main sequence are tracked but
are inert.  There is evidence of lower mass primordial stars forming
within relic \ion{H}{2} regions \citep{OShea05, Yoshida07b}, but we
neglect this to avoid additional uncertain parameters.  This is a
desired future improvement, however.

We use adaptive ray tracing \citep{Abel02b} to calculate the
photo-ionization and heating rates caused by stellar radiation.  We
consider photo-ionization from photons with an energy of 28.4 (29.2)
eV that is the mean energy of ionizing radiation from a metal-free
star with 100 (170) \Ms.  The simulation box is large enough so that
the \ion{H}{2} region never expands outside the box; however, our ray
tracing code employs isolated boundary conditions so that photons are
lost if they travel outside the computational domain.  We account for
\hh~photo-dissociation with a 1/$r^2$ Lyman-Werner radiation field
without self-shielding.  We use a non-equilibrium, nine-species (H,
H$^{\rm +}$, He, He$^{\rm +}$, He$^{\rm ++}$, e$^{\rm -}$, H$_2$,
H$_2^{\rm +}$, H$^{\rm -}$) chemistry solver in \enzo~\citep{Abel97,
  Anninos97} that takes into account the additional spatial dependence
of the photoionization rates provided by the radiation transport.

We end the simulations when the most massive halo begins to rapidly
collapse (i.e. \tcool~$<$ \tdyn) in the hydrogen and helium cooling
only runs at redshift 15.9 (16.8).  The virial temperature \tvir~of
the halo is $\sim$10$^4$ K at these redshifts.

%
%

\begin{deluxetable*}{cccccccc}
\tablecolumns{8}
\tabletypesize{}
\tablewidth{\textwidth}
\tablecaption{Selected Star Forming Halo Parameters\label{tab:halos}}

\tablehead{
  \colhead{\#} & \colhead{Sim} & \colhead{Type} & \colhead{$z$} &
  \colhead{M$_{{\rm vir}}$} & \colhead{$f_b$} & \colhead{$\rho_c$} &
  \colhead{$T_c$} \\
  \colhead{} & \colhead{} & \colhead{} & \colhead{} & \colhead{[\Ms]}
  & \colhead{} & \colhead{[cm$^{-3}$]} & \colhead{[K]}
} 
\startdata
 1 & SimB-RT & 1 & 30.9 & $4.7 \times 10^5$ & 0.081 & 1600 &  340 \\
 2 & SimA-RT & 1 & 29.9 & $4.8 \times 10^5$ & 0.094 & 6500 &  350 \\
 3 & SimB-RT & 4 & 23.7 & $5.3 \times 10^6$ & 0.059 & 2400 &  410 \\
 4 & SimB-SN & 4 & 21.0 & $1.1 \times 10^7$ & 0.045 & 1800 &  480 \\
 5 & SimA-RT & 4 & 20.4 & $6.3 \times 10^6$ & 0.069 &  550 &  440 \\
 6 & SimB-SN & 2 & 20.1 & $2.6 \times 10^6$ & 0.12  &  120 &  390 \\
 7 & SimB-RT & 2 & 19.9 & $2.8 \times 10^6$ & 0.12  &  870 &  450 \\
 8 & SimA-RT & 2 & 19.3 & $2.9 \times 10^6$ & 0.13  & 1300 &  440 \\
 9 & SimB-SN & 3 & 19.3 & $2.3 \times 10^6$ & 0.12  &  360 &  450 \\
10 & SimB-RT & 5 & 16.8 & $3.1 \times 10^7$ & 0.089 & 4100 & 2500 \\
11 & SimB-SN & 5 & 16.8 & $2.9 \times 10^7$ & 0.065 & 1100 &  590 \\
12 & SimA-RT & 5 & 16.1 & $3.0 \times 10^7$ & 0.061 &  130 &  470 
\enddata

\tablecomments{Col. (1): Halo number. Col. (2): Simulation
  source. Col. (3): Star formation type. Col. (4): Redshift. Col. (5):
  Virial mass.  Col. (6): Baryon mass fraction. Col. (7): Central
  number density. Col. (8): Central temperature.}

\end{deluxetable*}

\section{STAR FORMATION}
\label{sec:SF}

Here we describe the aspects of massive metal-free star formation in
our simulations.  The first star forms at redshift 29.7 (30.8) in halo
typical of Pop III star formation without any feedback that has a mass
of $\sim 5 \times 10^5 \Ms$ \citep[cf.][]{Abel00, Abel02, Machacek01,
  Yoshida03, Yoshida06}.  Afterwards there are a total of 19, 29, and
24 instances of star formation in SimA-RT, SimB-RT, and SimB-SN,
respectively.

%
%
\begin{figure}[b]
\begin{center}
\epsscale{1.15}
\plotone{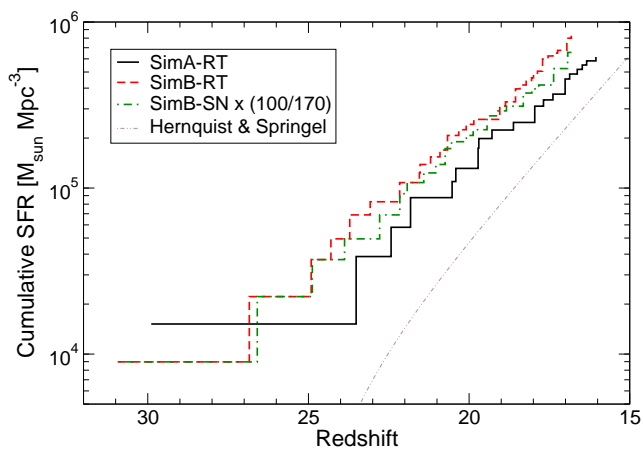}
\caption{\label{fig:cumulSF} Cumulative star formation rate in units
  of comoving \Ms~Mpc$^{-3}$ of SimA-RT (solid), SimB-RT (dashed), and
  SimB-SN (dot-dashed).  The star formation rate of SimB-SN has been
  scaled by 100/170, which is the ratio of Pop III stellar masses used
  in SimB-RT and SimB-SN, in order to make a direct comparison between
  the two simulations.  The dot-dot-dashed line represents the
  cumulative star formation rate in atomic hydrogen cooling halos from
  \citet{Hernquist03}.}
\end{center}
\end{figure}

\subsection{Star Formation Rate}

We show the cumulative star formation rate (SFR) in units of comoving
\Ms~Mpc$^{-3}$ in Figure \ref{fig:cumulSF}.  This quantity is simply
calculated by taking the total mass of stars formed at a given
redshift divided by the comoving volume where stars are allowed to
form (see \S\ref{sec:simulations}).  In this figure, we decrease the
SFR of SimB-SN by a factor of 1.7 in order to directly compare the
rates from the other two simulations.  This minimizes some of the
uncertainties entered into our calculations when we chose the free
parameter of Pop III stellar mass.  The cumulative rates are very
similar in both realizations.  The refined volume of Sim A (Sim B) has
an average overdensity $\delta \equiv \rho / \bar{\rho}$ = 1.4 (1.8).
The more biased regions in Sim B allows for a higher density of
star-forming halos that leads to the increased cumulative SFR.

We also overplot the cumulative SFR in atomic hydrogen cooling halos
from \citet{Hernquist03} in this figure.  It is up to an order of
magnitude lower than the rates seen in our calculations up to redshift
20.  They only focused on larger mass halos in their simulations.  The
disparity between the rates is caused by our simulations only sampling
a highly biased region, where we focus on a region containing a
3-$\sigma$ density fluctuation, and from the contribution from Pop III
stars.  The rates of \citet{Hernquist03} are calculated from an
extensive suite of smoothed particle hydrodynamics simulations that
encompasses both large and small simulation volumes and give a more
representative global SFR due to their larger sampled volumes.
However, our adaptive spatial resolution allows us to study both the
small- and large-scale radiative feedback from Pop III stars, which is
the main focus of the paper, in addition to the quantitative measures
such as a SFR.

%
%
\begin{figure}[b]
\begin{center}
\epsscale{1.15}
\plotone{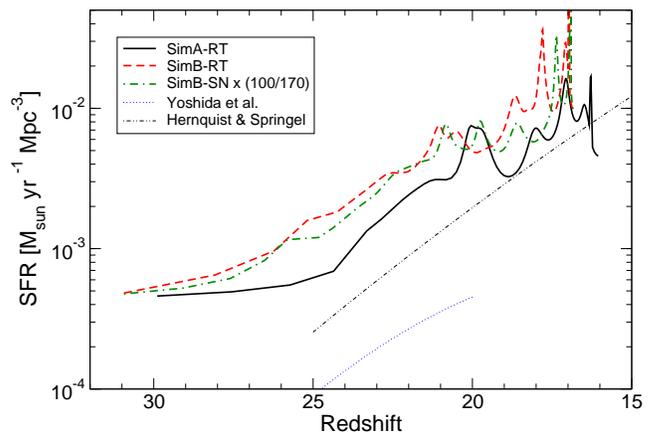}
\caption{\label{fig:SF} Comoving star formation rate in units of \sfr.
  The lines representing the simulation data have the same meaning as
  in Figure \ref{fig:cumulSF}.  The crosses show at which redshifts
  stars form.  The rates in SimB-SN are scaled for the same reason as
  in Figure \ref{fig:cumulSF}.  For comparison, we overplot the star
  formation rates from \citet{Hernquist03} in atomic hydrogen line
  cooling halos and \citet{Yoshida03} for 100\Ms~Pop III stars.}
\end{center}
\end{figure}

To estimate a volume-averaged SFR \citep[i.e.][in units of comoving
\sfr]{Madau96} from the cumulative SFR, we need to smooth the
discretely increasing cumulative SFR to ensure its time derivative
(i.e. the SFR) is a smoothly varying function.  Otherwise, the SFR
would be composed of delta functions when each star forms.  We first
fit the cumulative SFR with a cubic spline with 10 times the temporal
resolution.  Then we smooth the data back to its original time
resolution and evaluate its time derivative to obtain the SFR that we
show in Figure \ref{fig:SF}.  We also mark the redshifts of star
formation with crosses.  We again compare our rates to ones calculated
in \citet{Hernquist03} for metal-enriched stars and \citet{Yoshida03}
for Pop III stars with a mass of 100 \Ms.  Our rates are higher for
reasons discussed previously.  We do not advocate these SFRs as cosmic
averages but give them as a useful diagnostic of the performed
simulations.

We see an increasing function from $5 \times 10^{-4}$ \sfr~at redshift
30 to $\sim$$6 \times 10^{-3}$ \sfr~at redshift 20.  Here only one
star per halo forms in objects with masses $\lsim 5 \times 10^6 \Ms$.
Above this mass scale, star formation is no longer isolated in nature
and can be seen by the bursting nature of the star formation after
redshift 20, where the SFR fluctuates around $10^{-2}$
\sfr~\citep[cf.][]{Ricotti02b}.  Since we neglect \hh~self-shielding,
the strong Lyman-Werner (LW) radiation dissociates almost all \hh~in
the host halo and surrounding regions.  Thus we rarely see
simultaneous instances of star formation.  However, the regions that
were beginning to collapse when a nearby star ignites form a star 3 --
10 million years after the nearby star dies.  This only results in a
minor change in the timing of star formation.  Furthermore this delay
is minimal compared to the Hubble time and does not affect SFRs.

%
%

\begin{figure}[t]
\begin{center}
\epsscale{1.15}
\plotone{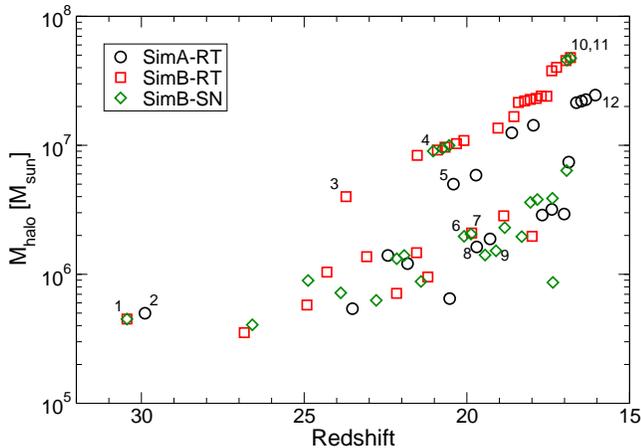}
\caption{\label{fig:SFhistory} Star formation times versus host halo
  DM mass for SimA-RT (circles), SimB-RT (squares), and SimB-SN
  (triangles).  One symbol represents one star.  The numbers
  correspond to the halo numbers listed in Table \ref{tab:halos}.}
\end{center}
\end{figure}

\subsection{Star Forming Halo Masses}
\label{sec:haloMass}


We show the star formation times versus the host halo DM masses as a
function of redshift in Figure \ref{fig:SFhistory}.  The DM halo
masses are calculated with the HOP algorithm \citep{Eisenstein98}.
First we focus on star formation in the largest halo.  Around redshift
30, the first star forms in all three simulations with a mass of
$\sim$$5 \times 10^5 \Ms$.  The stellar radiation drives a
$\sim$30\kms~shock wave that removes almost all of the gas from the
shallow potential well.  It takes approximately 75 (40) million years
for gas to reincorporate into the potential well from smooth IGM
accretion and mergers.  At $z \sim 24$ in SimB-RT, the second star
forms in the most massive progenitor that now has a mass of $4 \times
10^6 \Ms$.  In SimA-RT, the merger history is calmer at $z = 24-30$,
and enough gas becomes available for \hh~cooling and star formation at
$z \sim 20$.  Here the second star forms in the most massive
progenitor that has a mass of $5 \times 10^6 \Ms$.  In both RT
simulations, the stellar feedback expels most of the gas from its host
once again.  For Sim A (Sim B), another 10 (30) million years passes
before the next star to form in this halo.  Once the halo has a mass
of $\sim$10$^7$ \Ms, the potential energy is great enough to confine
most of the stellar and SNe outflows.  In SimA-RT and SimB-RT, halos
above this mass scale host multiple sites of star formation that is
seen in the nearly continuous bursts of star formation in the most
massive halo.  SimA-RT forms stars more intermittently than SimB-RT
because it undergoes two major mergers between redshift 17 and 21
\citep[see][]{Wise07a}.  In SimB-SN at $z = 21$, three stars form in
succession in the most massive halo.  Their aggregate stellar and SNe
feedback expels the gas from its halo one more time.  This halo only
forms another star at $z = 16.9$ (55 million years later) in this halo
when enough gas has been reincorporated.%
\footnote{We have run SimB-SN past $z = 16.8$ and have seen that it
  starts to host multiple sites of star formation.}

Most of the stars form in low-mass halos with masses $\sim$10$^6$
\Ms~that are forming its first star between redshifts 18 -- 25 in our
calculations.  A slight increase in host halo masses with respect to
redshift mainly occurs because of the negative feedback from
photo-evaporation of low-mass halos that are close to other
star-forming halos \citep{Haiman01}.  Additional delays in star
formation may be caused by ultraviolet heating and \hh~dissociation
from previous stars \citep[e.g.][]{Machacek01, Yoshida03, Mesigner06},
which increase the critical halo mass in which gas can cool and
condense.

One interesting difference in SimB-SN from the other calculations is
that star formation is sometimes induced in overdensities within the
same halo when a SN blast wave overtakes it.  This occurs in three
halos with masses of 2.0, 1.5, and 3.8 $\times 10^6 \Ms$ at redshifts
19.9, 19.1, and 17.8, respectively.  The same halos in SimB-RT do not
form two stars before their gas are expelled and thus quenching
subsequent star formation.

%
%
\begin{figure*}[p]
\begin{center}
\epsscale{1}
\plottwo{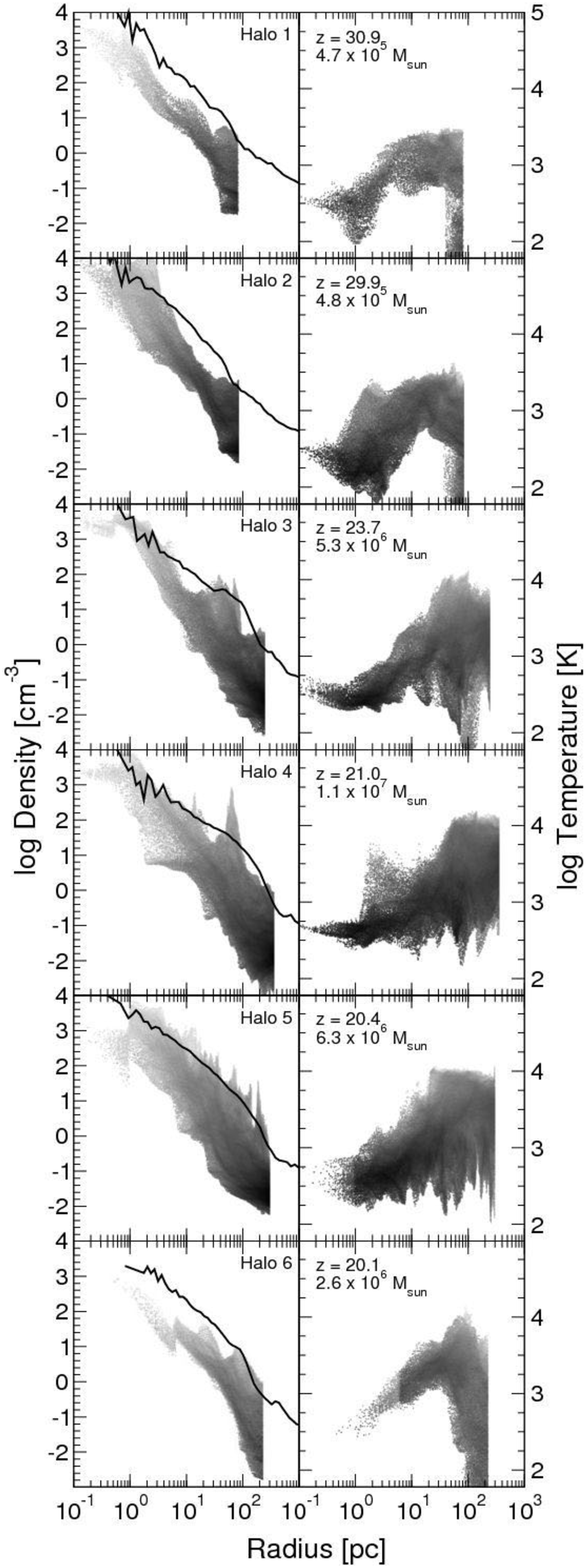}{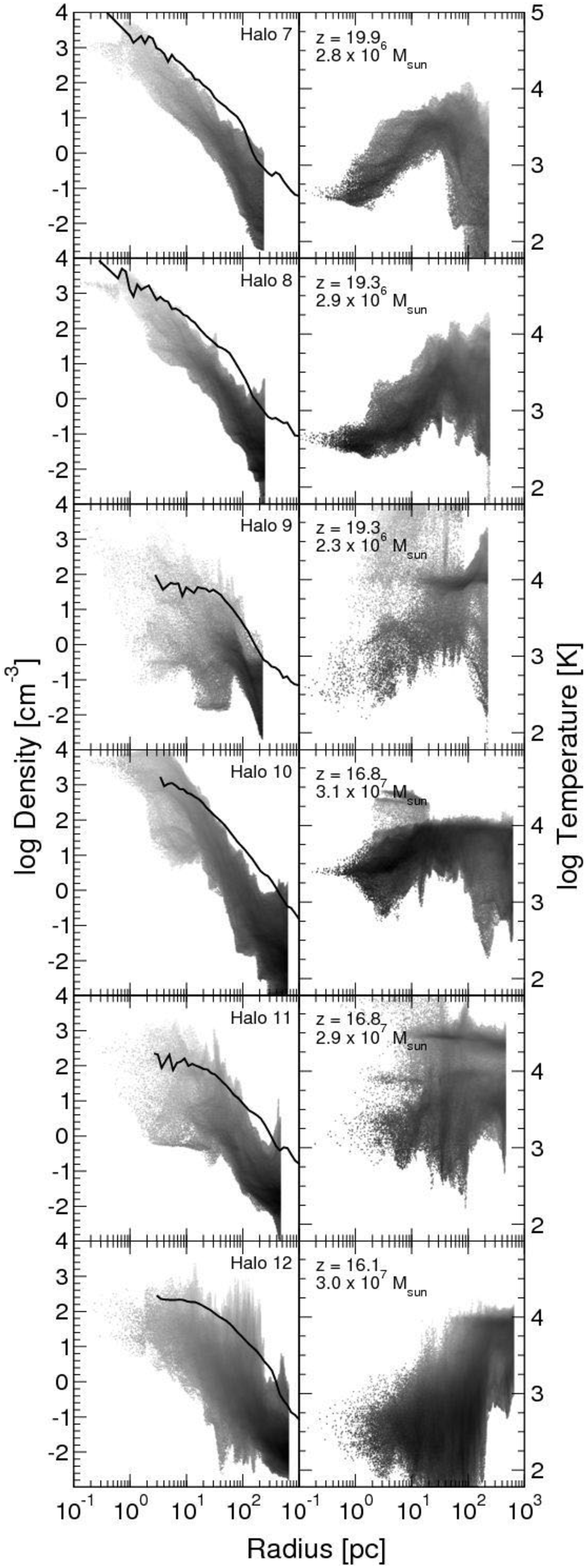}
\caption{\label{fig:radial} Radial profiles of number density (left
  column) and temperature (right column) for selected star forming
  halos inside the virial radius.  We overplot the radially averaged
  DM density (solid line) in units of $m_h$ \cubecm.  These data
  represent the state of the region immediately after star formation.
  Notice the added complexity (range) in the density and temperature
  with increasing host halo mass, especially if the region has been
  affected by stellar radiation, as in Halos 3, 4, 5, 10, 11, and 12.
  The occasional discontinuities in density in the inner parsec arise
  from our star formation recipe when we remove half the mass in a
  sphere containing twice the stellar mass.}
\end{center}
\end{figure*}

\subsection{Star Formation Environments}


We further study the nature of high-redshift star formation by
selecting four star forming regions from each simulation and studying
the surrounding interstellar medium (ISM) prior to star formation.
The ISM in the 10$^4$ K halos are described in more detail in
\citet{Wise07c}.  The sample of regions are chosen in order to compare
different star formation environments.  These regions can be
categorized into (1) first star inside an undisturbed halo, (2) first
star that is delayed by LW radiation, (3) induced star formation by
positive feedback, (4) star formation after gas reincorporation, and
(5) star formation in a halo with a virial temperature over 10$^4$ K.
The represented halos and their parameters are listed in Table
\ref{tab:halos} and annotated in Figure \ref{fig:SFhistory}.

We plot the mass-weighted radial profiles of number density (left
columns) and temperature (right columns) within the virial radius for
these twelve halos in Figure \ref{fig:radial} and describe them below.

\medskip

1. \textit{First star} (Halo 1, 2)--- These stars are the first to
form in their respective simulation volume.  The structure of the host
halos within our resolution limit exhibit similar characteristics,
e.g., a self-similar collapse and central temperatures of 300 K, as in
previous studies \citep{Abel00, Abel02, Bromm02, Yoshida06}.  The halo
masses are $4.8 (4.7) \times 10^5$ \Ms.  Heating from virialization
raises gas temperatures to 3000 K, and in the central parsec,
\hh~cooling becomes effective and cools the gas down to 200 K that
drives the further collapse.  The mass-weighted central gas densities
and temperatures are approximately 3000\cubecm~and 320 K,
respectively.

\medskip

2. \textit{Delayed first star} (Halo 6, 7, 8)--- The host halos have
similar radial profiles as the halos that hosted the first stars but
with masses of $\gsim 10^6 \Ms$.  Here the \hh~cooling has been
stifled by the LW radiation from nearby star formation.  Only when the
halo mass passes a critical mass, the core can cool and condense by
\hh~formation \citep{Machacek01, Yoshida03, OShea07, Wise07b}.  The
central densities are lower than the first stars with 1300, 870, and
120\cubecm~in SimA-RT, SimB-RT, and SimB-SN, respectively.  The
central temperatures are marginally higher at 440, 450, and 390 K.

\medskip

%
%

\begin{figure}[t]
  \begin{center}
    \epsscale{1.15}
    \plotone{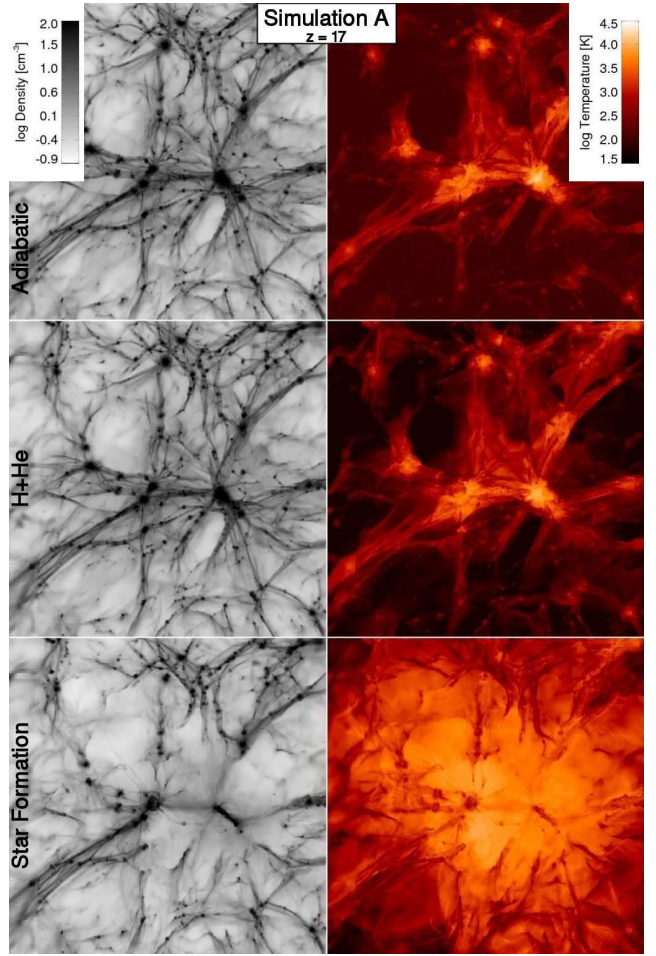}
    \caption{\label{fig:projA} Density-squared weighted projections of
      gas density (left) and temperature (right) of Sim A.  The field
      of view is 8.5 proper kpc (1/216 of the simulation volume) and
      the color scale is the same for all simulations.}
\end{center}
\end{figure}
%
%
\begin{figure}[t]
  \begin{center}
    \epsscale{1.15}
    \plotone{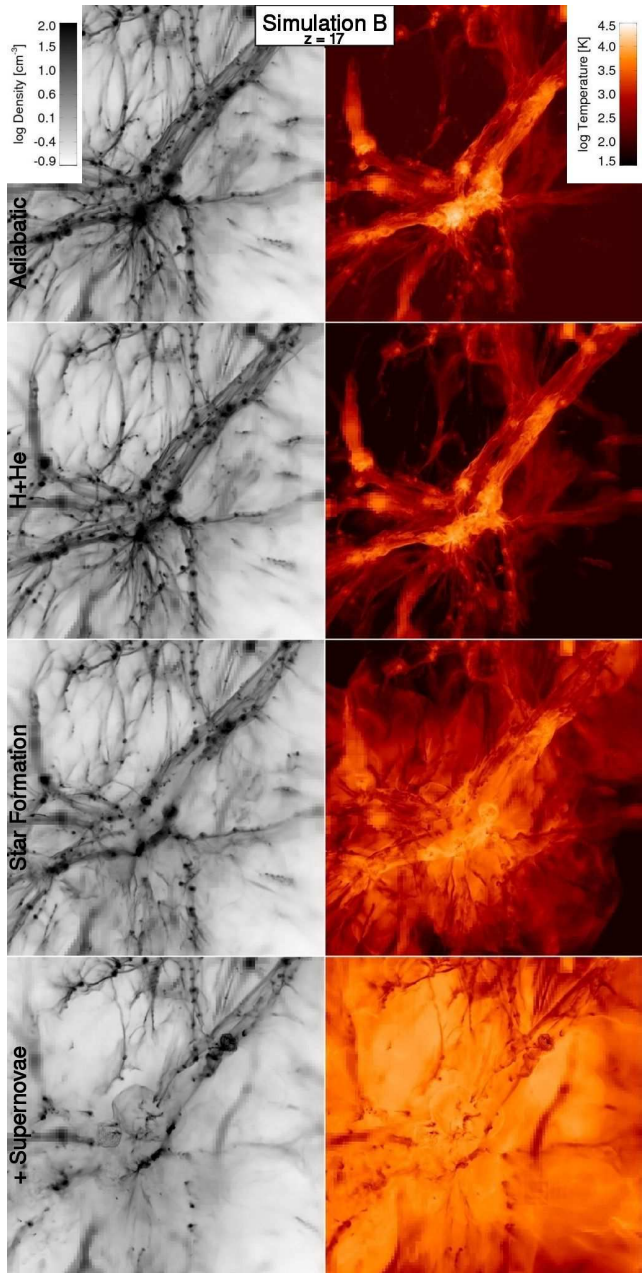}
    \caption{\label{fig:projB} Same as Figure \ref{fig:projA} but for
      Sim B.}
\end{center}
\end{figure}

3. \textit{Induced star formation} (Halo 9)--- At $z = 19.3$, a
massive star explodes in a SN, whose shell initially propagates
outward at 4000\kms.  After 160 kyr, the shell passes an overdensity
within the same halo that is caused by an ionization front instability
\citep[e.g., see][]{Whalen08a, Whalen08b}.  The star forms 35 pc away
from the SN explosion and the DM halo center.  The combination of the
shock passage and excess free electrons in the relic \ion{H}{2}
catalyze \hh~formation in this low-mass halo \citep[e.g.][]{Ferrara98,
  OShea05, Mesigner06}.  The SN blast wave heats the gas over 10$^4$ K
to radii as 1 pc.  In the density profile, both low and high density
gas exists at similar radii.  Here the shock passage creates a tail of
gas streaming from the central core, whose asymmetries can be seen in
the density profile.  However, the core survives and benefits from the
excess electrons created during this event.  The central temperature
is similar to the previous cases at 450 K.  The \hh~criterion for star
formation is reached faster because of the excess electrons, which
creates a star particle at a lower density (360\cubecm).

\medskip

4. \textit{Star formation after reincorporation} (Halo 3, 4, 5)---
After a sufficient amount of gas that was expelled by dynamical
feedback of the first star is reincorporated into the halo, star
formation is initiated again.  Here virial temperatures of the halos
are under 10$^4$~K, which are hosting their second instance of star
formation.  These halos have a larger spread in gas densities and
temperatures than the halos forming their first star.  Gas is heated
by virialization and prior stellar radiation to over 10$^4$ K outside
10 pc.  The central densities in halos 3 and 5 are similar to the
regions described in the first star formation section, however they
are slightly warmer at 410 and 480 K.  Halo 4 shows a more diffuse
core with densities of 550\cubecm~and temperatures of 440 K.

\medskip

5. \textit{Star formation in 10$^{\,4}$K halos} (Halo 10, 11, 12)---
In these halos, \hh~formation is aided by atomic hydrogen cooling.
The ISM becomes increasingly complex as more stars form in the halo.
The temperatures range from 100 K to 20,000 K throughout the halo.
Halos 10 and 12 have hosted 16 and 8 massive stars, respectively,
since it started to continually form stars at $z \sim 20$.  In SimB-RT
(halo 10), the densities are higher than the cases.  The gas in this
halo is more centrally concentrated than the others because the
\ion{H}{2} regions did not breakout of the halo, thus minimizing any
outflows from feedback and dispersion of the central core.  The
temperature in halo 10 is significantly warmer than other regions at
2500 K.  In halo 11, the devastation caused by three stars and their
SNe at $z = 21$ prevented star formation until $z = 16.9$.  Its
initial recovery from that event is apparent by the single cool core
with a temperature of 590 K that sharply transitions to a warm,
diffuse medium at $r = 10$ pc.  Halo 12 (SimA-RT) has a complex
morphology that is not centrally concentrated and is caused by stellar
outflows during a major merger \citep[see][for images]{Wise07c}.  This
morphology manifests itself in the radial profiles as large density
contrasts spanning nearly six order of magnitude at $r = 30-300$ pc.
Similarly, temperatures range from 50 K to 10,000 K in the same
region.  The star forms in a diffuse region ($\rho = 130\cubecm$) that
has a temperature of 470 K and whose \hh~formation is enhanced because
it resides in a relic \ion{H}{2} region.

\section{STARTING COSMOLOGICAL REIONIZATION}
\label{sec:reion}

In this section, we first describe the ionizing radiation from massive
stars that start cosmological reionization in small overdense regions
we simulate.  Then we discuss the effects of recombinations in the
inhomogeneous IGM and kinetic energy feedback from Pop III stars.
Lastly the evolution of the average IGM thermal energy is examined.

To illustratively demonstrate radiative feedback from massive stars on
the host halos and IGM, we show projections of gas density and
temperature that are density-squared weighted in Figures
\ref{fig:projA} and \ref{fig:projB} for all of the simulations at
redshift 17.  These projections have the same field of view of 8.5
proper kpc and the same color maps.  The large-scale density structure
is largely unchanged by the stellar feedback, and the adjacent
filaments remain cool since they are optically thick to the incident
radiation.  \hh~cooling produces more centrally concentrated objects;
however stellar feedback photo-evaporates $\lsim$10$^6$\Ms~halos near
other star-forming halos.  This is apparent in the density projections
in the Jeans smoothing around the most massive halo
\citep[cf.][]{Haiman01, Mesigner06}.  Kinematic feedback from SNe has
an even larger effect on the surrounding gas.  In SimB-SN, this effect
is seen in the reduced small-scale structure and low-mass halos with
no gas counterparts.  However, the most apparent difference in the
radiative simulations is the IGM heating by Pop III stars, especially
in SimB-SN.

%
%
\begin{figure}[b]
\begin{center}
  \epsscale{1.15}
\plotone{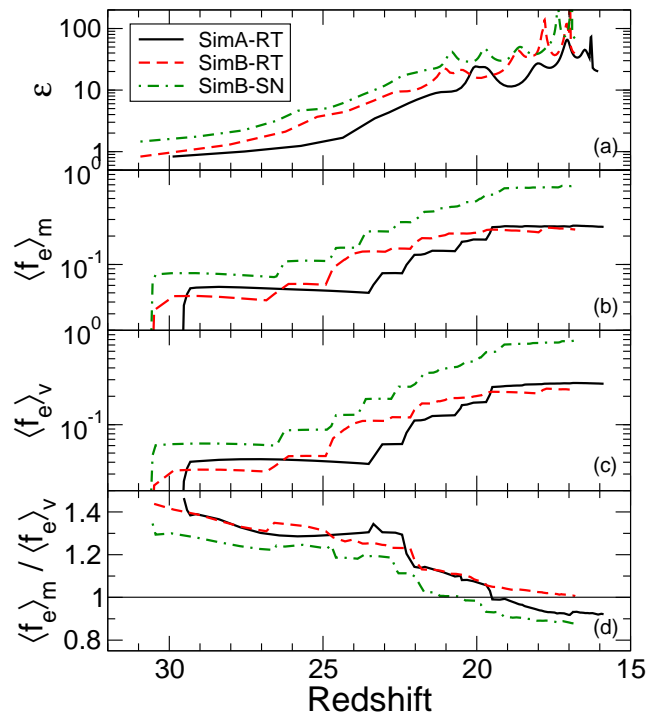}
\caption{\label{fig:emis} (a) Averaged emissivity in units of ionizing
  photons per baryon per Hubble time that is calculated from the star
  formation rate in Figure \ref{fig:SF}.  (b) Mass-averaged ionization
  fraction ($f_e > 10^{-3}$) of the inner 250 (300) comoving kpc for
  SimA (SimB).  (c) Volume-averaged ionization fraction for the same
  runs.  (d) The ratio of the mass- and volume-averaged ionization
  fraction.}
\end{center}
\end{figure}

\subsection{UV Emissivity}


A key quantity in reionization models is volume-averaged emissivity of
ionizing radiation.  We utilize the comoving SFR $\dot{\rho}_\star$ to
calculate the proper volume-averaged UV emissivity
\begin{equation}
  \label{eqn:emis}
  \epsilon = \frac{\dot{\rho}_\star Q_{\rm{HI}} t_{\rm{H}}}{\bar{\rho}_b}
\end{equation}
in units of ionizing photons per baryon per Hubble time.  Here
$Q_{\rm{HI}}$ is the number of ionizing photons emitted in the
lifetime of a star per solar mass, $\bar{\rho}_b \simeq 2 \times
10^{-7}$ is the comoving mean number density, and
\begin{equation}
  \label{eqn:hubbletime}
  t_{\rm{H}} \approx \frac{2}{3 H_0 \sqrt{\Omega_m}} (1+z)^{-3/2}
\end{equation}
is the Hubble time in a Einstein de-Sitter universe, which is valid
for $\Lambda$CDM cosmology at $z \gg 1$.  For a Pop III stellar masses
greater than 100 \Ms, $Q_{\rm{HI}} \approx 10^{62}$ photons per solar
mass, corresponding to 84000 ionizing photons per stellar proton
\citep{Schaerer02}.  We plot the emissivity $\epsilon$ in Figure
\ref{fig:emis}a.  It follows the same behavior as the SFR, but now can
be directly used in semi-analytic reionization models.  The emissivity
increases from unity at redshift 30 to $\sim$100 at the end of our
simulations.  Our results agree with the emissivity calculated in
semi-analytic models that include Pop III stars
\citep[e.g.][]{Onken04} and should be an upper limit however.

\subsection{Effective Number of Ionizations per UV Photon}

We show the mass-averaged and volume-averaged ionization fraction
$f_e$ within the refined region in Figures \ref{fig:emis}b and
\ref{fig:emis}c.  The ratio of these fractions are plotted in Figure
\ref{fig:emis}d.  The first star in the simulation ionizes between
5--10\% of the volume where we allow star formation.  As stars begin
to form in other halos after redshift 25, the ionization fraction
gradually builds to 30\% in the RT simulations and 75\% in the SN
case.  The higher stellar luminosities in SimB-SN, which can be seen
in Figure \ref{fig:emis}a, and the additional outflows generated by SN
blast waves cause this difference in $f_e$.  The ratio $\langle f_e
\rangle_m / \langle f_e \rangle_v$ illustrates that the ionized
regions are overdense by a factor of $\sim$1.25 at early times in the
simulation, but then decreases to unity as the \ion{H}{2} region
grows.  Additionally, the \ion{H}{2} regions in halos with sustained
star formation in the RT simulations do not fully breakout into the
IGM.  \citet{Kitayama04} provided a useful approximation of the
critical halo mass
\begin{equation}
  \label{eqn:ionCritMass}
  M_{\rm{crit}}^{\rm{ion}} \sim 2.5 \times 10^6
  \left(\frac{M_\star}{200 \Ms}\right)^{3/4}
  \left(\frac{1+z}{20}\right)^{-3/2} \Ms,
\end{equation}
in which an ionization front (I-front) cannot escape.  This
approximation is valid for stellar masses between 80 and 500 \Ms,
redshifts between 10 and 30, and singular isothermal spheres.  Our
simulations exhibit this same trait in which I-fronts only partially
breakout from the host halo above this mass scale.

%
%
\begin{figure}[t]
\begin{center}
  \epsscale{1.15}
\plotone{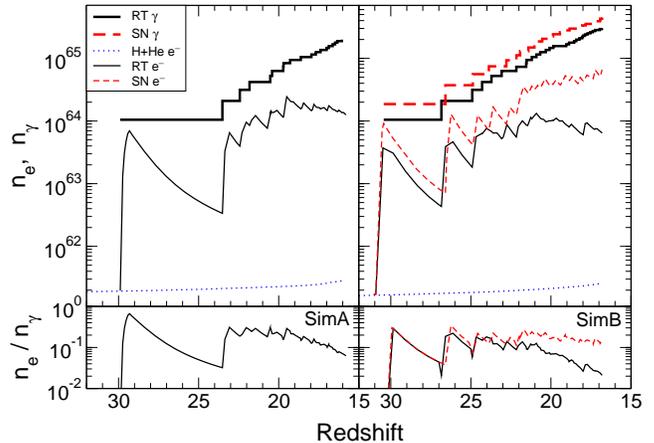}
\caption{\label{fig:elec} \textit{Top panels}: Total number of
  ionizing photons emitted (thick lines) and total number of electrons
  (thin lines) for simulations with cooling only (dotted), star
  formation only (solid), and supernovae (dashed) in the inner 250 and
  300 comoving kpc for SimA (left) and SimB (right).  The \ion{H}{2}
  regions are completely contained in these volumes.  \textit{Bottom
    panels}: The ratio of total number of electrons to the total
  number of ionized photons emitted.}
\end{center}
\end{figure}

This is not the case with SNe because previous SN blast waves can more
effectively evacuate the surrounding medium, thus increasing the
chances of radiation escaping into the IGM from later stars in the
same halo.  At $z = 21$, there is an example of this occurring with
three stars forming in succession in the most massive halo.  After the
first star goes SN, a diffuse and hot medium is left behind, but the
blastwave has not completely disrupted two other nearby condensing
clumps.  The radiation from the second star now does not have to
ionize its host halo and has an escape fraction of near unity.  The
same happens for the third star in this halo.  This episode further
ionizing SimB-SN from 40\% to 60\%.  As a note of caution, these
ionized fractions should not be considered as cosmological average
because they only sample a highly biased region.  \citet{Iliev06}
showed that a simulation box size of $\sim$30 Mpc is needed to make
global predictions.

To examine the strength of recombinations, we compare the total number
of electrons in the volume to the total number of ionizing photons
emitted in Figure \ref{fig:elec}.  The ratio of these two quantities
is the number of UV photons needed for one effective ionization
initially.  This ratio is approximately 3/5 (1/3) after the first star
dies.  The values in simulation A are higher due to its smaller
volume.  This ratio then steadily decreases from recombinations in the
relic \ion{H}{2} region to a few percent when the next star forms.  As
stars begin to form regularly in the simulation, there are 4 (6)
photons per sustained ionization.  However this ratio drops by a
factor of 5 in the RT simulations after $z \sim 20$ when the
\ion{H}{2} regions become trapped in the halo, thus reducing the
available photons for ionizing the IGM.  The effects of SNe as
previously discussed maintains the ratio of 6 photons required per
ionization as the most massive halo grows in mass.

%
%
\begin{figure}[t]
\begin{center}
  \epsscale{1.15}
\plotone{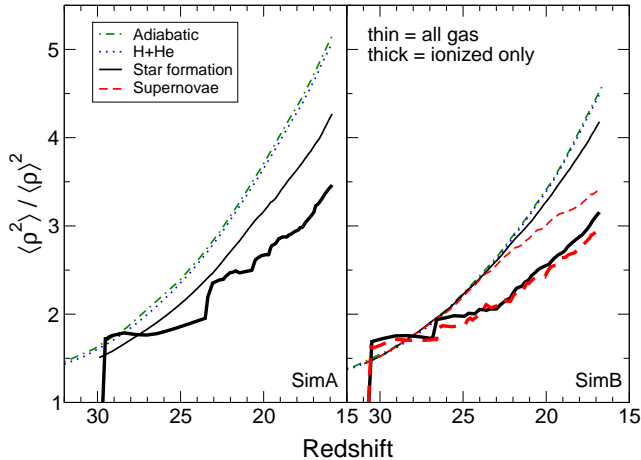}
\caption{\label{fig:clumping} Clumping factor $C =
  \langle\rho^2\rangle / \langle\rho\rangle^2$ for SimA (left) and
  SimB (right), comparing the cases of the adiabatic equation of state
  (dot-dashed), atomic hydrogen and helium cooling (dotted), star
  formation only (solid), and supernovae (dashed).  The clumping
  factors for all diffuse ($\delta < 100$) gas are plotted as thin
  lines, and the clumping factors for ionized, diffuse gas,
  $C_{\rm{ion}}$, are plotted as thick lines.}
\end{center}
\end{figure}

\subsection{Clumping Factor Evolution}

Volume averaged recombination rates in an inhomogeneous IGM scale with
the clumping factor $C = \langle\rho^2\rangle / \langle\rho\rangle^2$,
where the angled brackets denote volume averaged quantities.  The
recombination rate for hydrogen, e.g., is simply
\begin{equation}
  \label{eqn:clumping}
  \left(\frac{dn_{\rm{HII}}}{dt}\right)_{\rm{rec}} = C k_{\rm{rec}}
  f_e \bar{\rho}_b (1+z)^3 ,
\end{equation}
where $k_{\rm{rec}}$ is the case B recombination rate for hydrogen at
$T \approx 10^4$ K, and $f_e = n_e/n$ is the ionization fraction.
Both the increased recombinations in overdense regions and photon
escape fractions lower than unity result in the high number of UV
photons needed for one effective ionization that we see in our
simulations.

Figure \ref{fig:clumping} compares the clumping factor in the
adiabatic, cooling only, star formation, and supernovae calculations,
which we separate into $C$ for all gas (thin lines) and ionized gas
(thick lines).  Since we resolve the local Jeans length by at least 4
cells in all simulations, the clumping factor is not underestimated,
given our assumptions about gas cooling in each model.  Recall that
our simulations use the Zel'dovich approximation for the initial
conditions, which may lead to underestimating the clumping factor.  We
do not include dense ($\delta > 100$) gas, residing in filaments and
halos, in the clumping factor calculation as we restrict this analysis
to the IGM because these overdensities are self-shielded from incident
ionizing radiation \citep[e.g.][]{Shapiro97, Shapiro04, Iliev05,
  Whalen08c}.

The clumping factor in the adiabatic and cooling-only cases smoothly
increases to $\sim$5 at $z = 17$ from unity at $z > 30$.  The clumping
factor in the star formation and supernovae simulations are smaller by
10--25\% than the other simulations because the overdensities in the
IGM are photoevaporated by the nearby stars.  SN explosions disperse
gas more effectively than radiative feedback alone in larger halos and
can have a bigger impact on the clumping factor.  At redshift 20, the
three stars and their SNe energy in the most massive halo destroy the
surrounding baryonic structures and reduce the clumping back to the
values seen in non-radiative cases.

We plot the clumping factor $C_{\rm{ion}}$ in ionized regions above
$x_e > 10^{-3}$ as the thick lines in Figure \ref{fig:clumping}.  This
value is most relevant for recombination rates.  Around the first star
in the simulation, the IGM is overdense and contains a larger amounts
of clumps than the rest of the simulation.  This increased clumpiness
competes with photoevaporation caused by the nearby star to create
$C_{\rm{ion}}$ values that are comparable to $C$ for all gas at $z >
25$.  Afterwards $C_{\rm{ion}}$ is always smaller than $C$ with a
maximum decrease of $\sim$25\%.


\subsection{Kinetic Energy Feedback}
\label{sec:kinetic}

SN explosion energy and kinetic energy generated in D-type I-front
play a key role in star formation in low-mass halos, which are easily
affected due to their shallow potential well \citep[e.g.][] {Dekel86,
  Haehnelt95, Bromm03, Whalen04, Kitayama04, Kitayama05}.  The kinetic
energy created by SNe is sufficient to expel the gas from these
low-mass halos.  For example, the binding energy of a $10^6 \Ms$ halo
is only $2.8 \times 10^{50}$ erg at $z = 20$, which is two orders of
magnitude smaller than a typical energy output of a pair instability
SN \citep{Heger02}.  For a \tvir~$>$ 10$^4$ K halo at the same
redshift, it is $9.4 \times 10^{52}$ erg.  With our chosen stellar
mass of 170 \Ms, it takes 3 -- 4 SNe to overcome this potential
energy.

The shock wave created by the D-type I-front travels at a velocity
$v_s$ = 25 -- 35\kms~for density gradients (i.e. $\rho(r) \propto
r^{-w}$) with slopes between 1.5 and 2.25 \citep{Shu02, Whalen04,
  Kitayama04}.  This velocity is the escape velocity for halos with
masses greater than $3 \times 10^8 \Ms$ at $z = 15$, which is an order
of magnitude greater than the most massive halos studied here.
However less massive halos can contain these I-fronts because pressure
forces slow the I-front after the star dies.  

Using the position of the shock wave when the star dies
(eq. \ref{eqn:shockPos}) and energy arguments, we can estimate the
critical halo mass where the material in the D-type I-front can escape
from the halo by comparing the binding energy $E_b$ of the halo and
kinetic energy in the shell.  For most massive stars, the shock wave
never reaches the final Str{\"o}mgren radius,
\begin{equation}
  \label{eqn:stroemgren}
  R_{\rm{str}} = 150 \left(\frac{\dot{N}_{\rm{HI}}}{10^{50} \mathrm{ph \;
        s}^{-1}}\right)^{1/3}
  \left(\frac{n_f}{1\cubecm}\right)^{-2/3} \; \mathrm{pc} ,
\end{equation}
before the star dies.  Here $\dot{N}_{\rm{HI}}$ is the ionizing photon
rate of the star, and $n_f$ is the average number density of gas
contained in this radius.  After the lifetime of the star, the shock
reaches a radius
\begin{equation}
  \label{eqn:shockPos}
  R_s = 83 \left(\frac{v_s}{30\kms}\right)
  \left(\frac{t_\star}{2.7 \mathrm{Myr}}\right) \; \mathrm{pc} ,
\end{equation}
where $t_\star$ is the stellar lifetime \citep[see
also][]{Kitayama04}.  We can neglect isolated, lower mass (M $\lsim$
30\Ms) Pop III stars whose shock wave reaches $R_{\rm{str}}$ within
its lifetime.  In this case, the I-front stops at $R_{\rm{str}}$, and
the shock wave becomes a pressure wave that has no associated density
contrast in the neutral medium \citep{Shu92}.  Thus we can safely
ignore these stars in this estimate.


Assume that the source is embedded in a single isothermal sphere.  The
mass contained in the shell is
\begin{equation}
  \label{eqn:massswept}
  M_{\rm{sw}} = \frac{(\Omega_b/\Omega_M) \> M_{\rm{vir}} \>
    R_s}{r_{\rm{vir}}} - V_s \rho_i
\end{equation}
that is the mass enclosed in the radius $R_s$ in an isothermal sphere,
corrected for the warm, ionized medium behind the I-front.  Here $V_s$
is the volume contained in a sphere of radius $R_s$, and $\rho_i$ is
the gas density of the ionized medium, whose typical number density is
1\cubecm~for stellar feedback from a massive primordial star
\citep{Whalen04, Kitayama04, Yoshida07a, Abel07}.  For massive stars
($M_\star \gsim 30 \Ms$), the mass of the central homogeneous medium
is small (i.e. 10\%) compared to the shell.  We compensate for this
interior mass by introducing the fraction $\eta$, so the shell mass is
simply
\begin{equation}
  \label{eqn:swept2}
  M_{\rm{sw}} = \eta \; \frac{(\Omega_b/\Omega_M) \> M_{\rm{vir}} \>
    r_s}{r_{\rm{vir}}}.
\end{equation}
For these outflows to escape from the halo, the kinetic energy
contained in the shell must be larger than the binding energy, which
is
\begin{equation}
  \label{eqn:outflowE}
  \frac{1}{2} M_{\rm{sw}} v_s^2 \; > \; \frac{G M_{\rm{vir}}^2}{2 r_{\rm{vir}}}.
\end{equation}
Using equations (\ref{eqn:shockPos}) and (\ref{eqn:swept2}) in this
condition, we obtain the maximum mass
\begin{eqnarray}
  \label{eqn:criticalMass}
  M_{\rm{max}} &\; \sim \;& \frac{r_s \> v_s^2 \> \Omega_b}{G \>
    \Omega_M} \nonumber\\
  M_{\rm{max}} & \sim & 3.20 \times 10^6 \left(\frac{r_s}{100 \mathrm{pc}}\right)
  \left(\frac{v_s}{30 \kms}\right)^2 \nonumber\\
  &   & \times \left(\frac{\eta}{0.9}\right)
  \left(\frac{\Omega_b/\Omega_M}{0.17}\right) \Ms
\end{eqnarray}
of a halo where the material in the shock wave becomes unbound,
expelling the majority of the gas from the halo.

%
%
\begin{figure}[t]
\begin{center}
  \epsscale{1.15}
\plotone{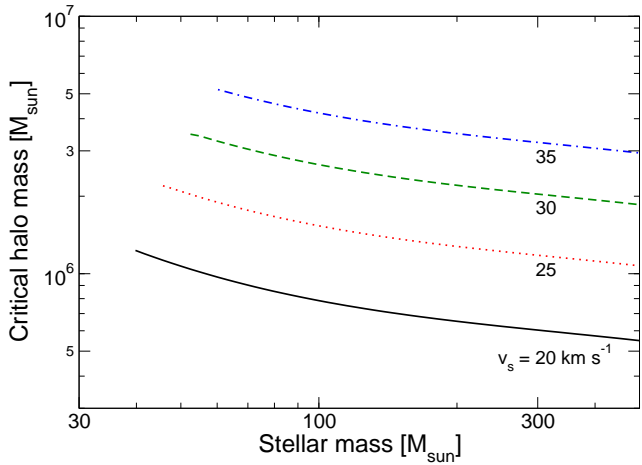}
\caption{\label{fig:critMass} Maximum halo mass in which a D-type
  ionization front can create outflows as a function of primordial
  stellar mass for shock velocities $v_s$ of 20, 25, 30, and 35\kms.
  Here the fraction $\eta$ of mass contained in the shell is 0.9.}
\end{center}
\end{figure}

In Figure \ref{fig:critMass}, we use the stellar lifetimes and
ionizing luminosities from \citet{Schaerer02} to calculate the
critical halo mass for outflows for stellar masses 5 -- 500\Ms~and for
shock velocities of 20, 25, 30, and 35\kms~with $\eta = 0.9$.  For
stellar masses smaller than 30\Ms, the D-type I-front reaches the
final Str{\"o}mgren sphere and cannot expel any material from the
host.  Hence they are not plotted in this figure.  For the more
massive stars, the star dies before the D-type I-front can reach the
Str{\"o}mgren radius, thus being limited by $t_\star$.  This maximum
halo mass is in good agreement with our simulations as we see halos
with masses greater than $5 \times 10^6\Ms$ retaining most of their
gas in the star formation only cases.  However in larger halos,
stellar sources still generate champagne flows, but this material is
still bound to the halo and returns in tens of million years.

%
%
\begin{figure}[b]
\begin{center}
  \epsscale{1.15}
\plotone{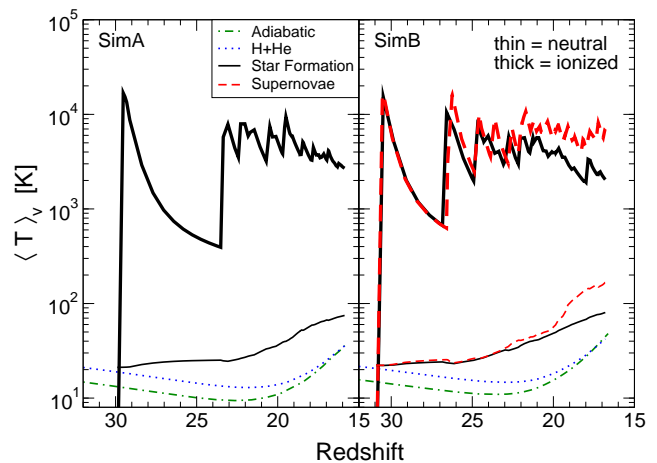}
\caption{\label{fig:volTemp} Evolution of the volume-averaged
  temperature in the inner 250 and 300 comoving kpc for Sim A (left)
  and Sim B (right), respectively, for neutral (thin) and ionized
  (thick; $f_e > 10^{-3}$) gas.  The simulations for the adiabatic
  (dot-dashed), cooling only (dotted), star formation only (solid),
  and supernovae (dashed) simulations are plotted.}
\end{center}
\end{figure}

\subsection{Thermal Energy}

Thermal feedback is yet another mechanism how Pop III stars leave
their imprint on the universe.  The initial heating of the IGM will
continue and intensify from higher SFRs at lower redshifts
\citep[e.g.][]{Hernquist03, Onken04}.  It is possible to constrain the
reionization history by comparing temperatures in the \lya~forest to
different reionization scenarios \citep{Hui03}.  Temperatures in the
\lya~forest are approximately 20,000~K at $z = 3 - 5$ \citep{Schaye00,
  Zaldarriaga01}.  Although our focus was not on redshifts below 15
due to the uncertainty of the transition to the first low-mass
metal-enriched (Pop II) stars, we can utilize the thermal data in our
radiation hydrodynamical simulations to infer the thermal history of
the IGM at lower redshifts.

The excess energy from hydrogen ionizing photons over 13.6 eV
photo-heat the gas in the \ion{H}{2} region.  The mean temperature
within \ion{H}{2} regions in our calculations is $\sim$30,000 K.  When
the short lifetime of a Pop III star is over, the \ion{H}{2} region
cools mainly through Compton cooling off the cosmic microwave
background.  The same framework applies to SNe remnants as well.  The
timescale for Compton cooling is
\begin{equation}
  \label{eqn:compton}
  t_C = 1.4 \times 10^7 \left(\frac{1+z}{20}\right)^{-4} f_e^{-1} \; \mathrm{yr}.
\end{equation}
This process continues until the gas recombines, and Compton cooling
is no longer efficient because of its dependence on electron fraction.
Radiation preferentially propagates into the voids and leaves the
adjacent filaments and its embedded halos virtually untouched.  Hence
we can restrict the importance of Compton cooling to the diffuse IGM
since Compton cooling cools the gas to low temperatures without being
impeded by recombinations that are proportional to $n_e^2$.  This
causes the relic \ion{H}{2} region to cool to temperatures down to
300~K.  The temperature evolution in our radiative calculations agrees
with the analytic models of relic \ion{H}{2} \citep{Oh03}.

%
%
\begin{figure}[b]
  \begin{center}
  \epsscale{1.15}
    \plotone{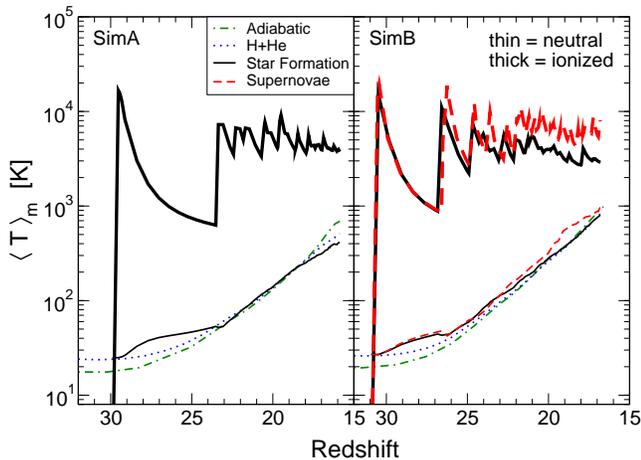}
    \caption{\label{fig:massTemp} Same as Figure \ref{fig:volTemp} but
      for the mass-averaged temperature.}
\end{center}
\end{figure}

We plot the volume-averaged temperature \tv~and mass-averaged
temperature \tm~in the volume where we allow star formation, i.e. the
inner 250 (300) comoving kpc, in Figures \ref{fig:volTemp} and
\ref{fig:massTemp}.  We compute the average temperatures in both
neutral and ionized ($f_e > 10^{-3}$) regions.  We first focus on the
thermal evolution of neutral gas.  An increase of the average
temperature in neutral gas is indicative of heating by hard photons or
supernovae.  At the final redshift $z = 16.8$, \tv~= 180 K in neutral
gas with SNe compared with 90 K without SNe.  Both of these average
temperatures are a factor of 2--3 higher than without star formation.
With SNe, \tm~also increases by $\sim$100 K to 1000 K.  In the SimA
panel of Figure \ref{fig:massTemp}, radiative cooling in neutral gas
by \lya~and \hh~in SimA-HHe and SimA-RT, respectively, is apparent at
the final redshift.  In the star formation run, most of the neutral
mass fraction lie within the most massive halo and are shielded from
radiation from stars within the halo and thus can radiatively cool.
This occurs in SimB but to a lesser extent and is not clearly seen in
Figure \ref{fig:massTemp}.

The effects from radiative feedback is most evident in the average
temperatures of ionized gas.  The first star in the calculations
raises the mass- and volume-averaged temperatures of the ionized gas
to $2 \times 10^4$~K.  Afterwards the remnant cools from Compton and
adiabatic processes as it expands to temperatures similar to the RT
simulations.  Photoheating from later stars cause the temperatures in
the ionized regions to fluctuate between 3000 and 10000 K.  The
supernovae calculations are slightly higher due to the hot SN bubble
that has an initial temperature of $\sim$$10^8$~K.  The mass-averaged
temperature increases more than \tv~because of the photo-heating of
the host halo and virial heating of the halos, which is the cause of
the increase in the simulations without star formation.  These
increased temperatures cause the photo-evaporation and Jeans smoothing
of the gas in the relic \ion{H}{2} regions.  We discuss these effects
in the next section.





\section{DISCUSSION}
\label{sec:discussion}

We have studied the details of massive metal-free star formation and
its role in the start of cosmological reionization.  We have treated
star formation and radiation in a self-consistent manner, allowing for
an accurate investigation of the evolution of cosmic structure under
the influence of early Pop III stars.  Stellar radiation from these
stars provides thermal, dynamical, and ionizing feedback to the host
halos and IGM.  Although Pop III stars are not thought to provide the
majority of ionizing photons needed for cosmological reionization,
they play a key role in the early universe because early galaxies that
form in these relic \ion{H}{2} regions are significantly affected by
Pop III feedback.  Hence it is important to consider primordial
stellar feedback while studying early galaxy formation.  In this
section, we compare our results to previous numerical simulations and
semi-analytic models of reionization and then discuss any potential
caveats of our methods and possible future directions of this line of
research.

%
%

\begin{figure}[t]
  \begin{center}
  \epsscale{1.15}
    \plotone{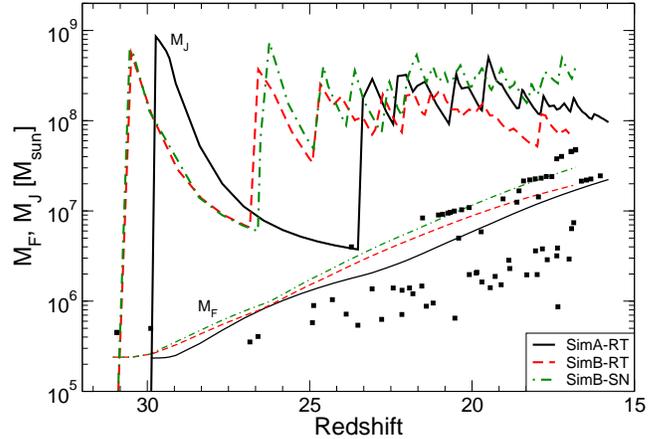}
    \caption{\label{fig:filtering} The Jeans mass $M_J$ and filtering
      mass $M_F$ that can form bound objects.  The squares denote the
      total mass of star forming halos in all three simulations.}
\end{center}
\end{figure}

\subsection{Comparison to Previous Models}

\subsubsection{Filtering Mass}
\label{sec:filtering}

One source of negative feedback is the suppression of gas accretion
into potential wells when the IGM is preheated.  The lower limit of
the mass of a star forming halo is the Jeans filtering mass
\begin{equation}
  \label{eqn:filtering}
  M_F^{2/3}(a) = \frac{3}{a} \int_0^a \> da^\prime
  M_J^{2/3}(a^\prime) \left[ 1 - \left(\frac{a^\prime}{a}\right)
  \right],
\end{equation}
where $a$ and $M_J$ are the scale factor and time dependent Jeans mass
in the \ion{H}{2} region \citep{Gnedin98, Gnedin00b}.  Additionally,
the virial shocks are weakened if the accreting gas is preheated and
will reduce the collisional ionization in halos with $\tvir \gsim
10^4$ K.  To illustrate the effect of Jeans smoothing, we take the
large \ion{H}{2} region of SimB-SN because it has the largest ionized
filling fraction, which is constantly being heated after $z = 21$.
Temperatures in this region fluctuates between 1,000~K and 30,000~K,
depending on the proximity of the currently living stars.  In Figure
\ref{fig:filtering}, we show the resulting filtering mass of regions
with an ionization fraction greater than $10^{-3}$ along with the
total mass of star forming halos.

\citet{Gnedin00b} found the minimum mass of a star forming halo is
better described by $M_F$ instead of $M_J$.  Our simulations are in
excellent agreement for halos that are experiencing star formation
after reincorporation of their previously expelled gas.  The filtering
mass is the appropriate choice for a minimum mass in this case as the
halo forms from preheated gas.  However for halos that have already
assembled before they become embedded in a relic \ion{H}{2} region,
the appropriate minimum mass $M_{\rm{min}}$ is one that is regulated
by the LW background \citep{Machacek01, Wise05} and photo-evaporation
\citep[e.g.][]{Efstathiou92, Barkana99, Haiman01, Mesigner06}.  This
is evident in the multitude of star forming halos below $M_F$.  With
the exception of star formation induced by SN blast waves or I-fronts,
this verifies the justification of using $M_{\rm{min}}$ and $M_F$ for
Pop III and galaxy formation, respectively, as a criterion for star
forming halos in semi-analytic models.

%
%
\begin{figure*}[t]
  \begin{center}
  \epsscale{1.15}
    \plotone{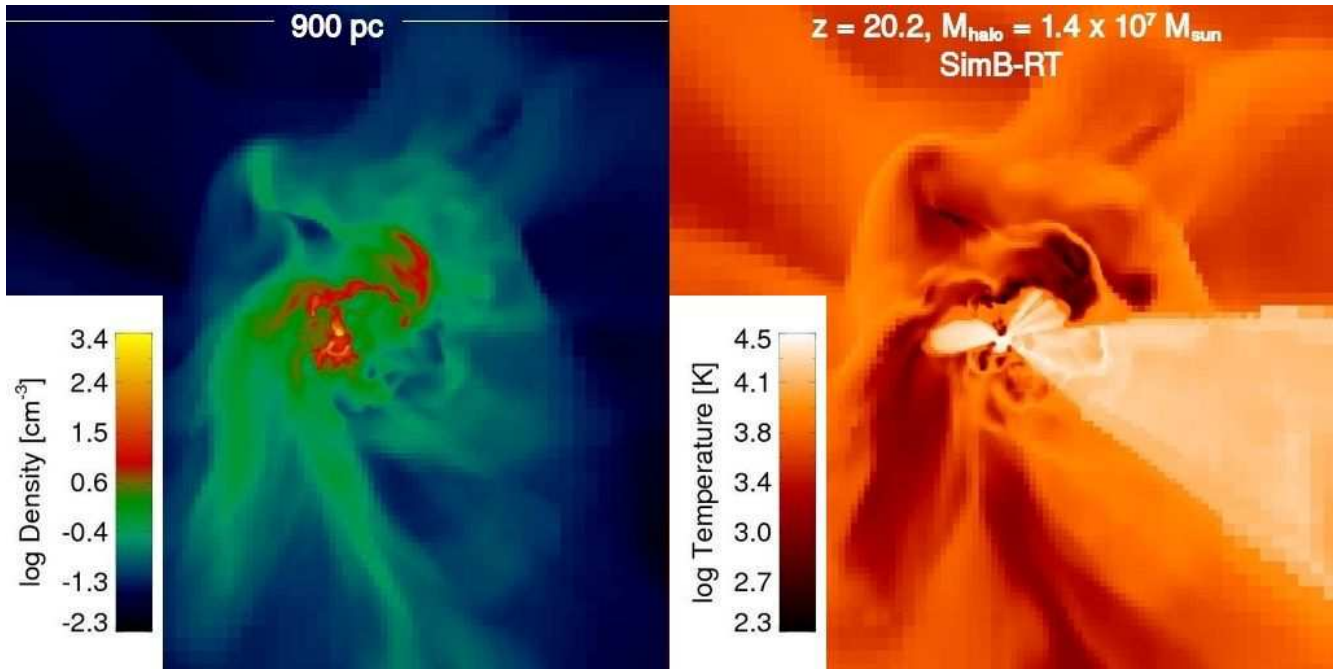}
    \caption{\label{fig:aniso} Density (left) and temperature (right)
      slices of an anisotropic \ion{H}{2} region in the most massive
      halo of SimB-RT.  The star has lived for 2.5 Myr out of its 2.7
      Myr lifetime.  The field of view is 900 proper parsecs.}
  \end{center}
\end{figure*}


\subsubsection{Star Formation Efficiency}
\label{sec:SFeff}

Semi-analytic models rely on a star formation efficiency $f_\star$,
which is the fraction of collapsed gas that forms stars, to calculate
quantities such as emissivities, chemical enrichment, and IGM
temperatures.  Low-mass halos that form a central star have $f_\star
\sim 10^{-3}$ whose value originates from a single 100~\Ms~star
forming in a dark matter halo of mass 10$^6$~\Ms~\citep{Abel02,
  Bromm02, Yoshida06}.  Pop II star forming halos are usually
calibrated with star formation efficiencies from local dwarf and
high-redshift starburst galaxies and are usually on the order of a few
percent \citep[e.g.][]{Taylor99, Gnedin00a}.

This leads to the question: how efficient is star formation in these
high-redshift halos while explicitly considering feedback?  This is
especially important when halos start to form multiple massive stars
and when metallicities are not sufficient to induce Pop II star
formation.  The critical metallicity for a transition to Pop II is
still unclear.  Recently, \citet{Jappsen07a} showed that metal line
cooling is dynamically unimportant in diffuse gas until metallicities
of $10^{-2} \> Z_\odot$.  On the other hand, dust that is produced in
SNe can generate efficient cooling down in dense gas with $10^{-6} \>
Z_\odot$ \citep{Schneider06}.  If the progenitors of the more massive
halos did not result in a pair-instability SN, massive star formation
can continue until it becomes sufficiently enriched.  Hence our
simulations can probe the efficiency of this scenario of massive
metal-free star formation.  It has also been suggested that the
cosmological conditions that lead to the collapse of a metal-poor
molecular cloud ($Z/Z_\odot \approx 10^{-3.5}$) may be more important
than some critical metallicity in determining the initial mass
function of a given stellar system \citep{Jappsen07b}.

We calculate $f_\star$ with the ratio of the sum of the stellar masses
to the total gas mass of unique star-forming halos.  For example at
the final redshift of 15.9 in SimA-RT, the most massive halo and its
progenitors had hosted 11 stars and the gas mass of this halo is $1.8
\times 10^6 \Ms$, which results in $f_\star = 6.1 \times 10^{-4}$ for
this particular halo.  Expanding this quantity to all star forming
halos, $f_\star/10^{-4} = 5.6, 6.7, 7.4$ for SimA-RT, SimB-RT, and
SimB-SN, respectively.  We note that our choice of $M_\star = 170 \Ms$
in SimB-SN increases $f_\star$ by 70\%.  Our efficiencies are smaller
than the isolated Pop III case because halos cannot form any stars
once the first star expels the gas, and 40 -- 75 million years must
pass until star can form again when the gas is reincorporated into the
halo.

By regarding the feedback created by Pop III stars and associated
complexities during the assembly of these halos, the $f_\star$ values
of $\sim$$6 \times 10^{-4}$ that are explicitly determined from our
radiation hydrodynamical simulations provide a more accurate estimate
on the early star formation efficiencies.

\subsubsection{Intermittent \& Anisotropic Sources}

Our treatment of star formation and feedback produces intermittent
star formation, especially in low-mass halos.  If one does not account
for this, star formation rates might be overestimated in this phase of
star formation.  Kinetic energy feedback is the main cause of this
behavior.  As discussed in sections \ref{sec:haloMass} and
\ref{sec:kinetic}, shock waves created by D-type I-fronts and SN
explosions expel most of the gas in halos with masses $\lsim 10^7$
\Ms.  A period of quiescence follows these instances of star
formation.  Then stars are able to form after enough material has
accreted back into the halo.  Only when the halo becomes massive
enough to retain most of the outflows and cool efficiently through
\lya~and \hh~radiative processes, star formation becomes more regular
with successive stars forming.

The central gas structures in the host halo are usually anisotropic as
it is acquiring material through accretion along filaments and
mergers.  At scales smaller than 10 pc, the most optically thick
regions produce shadows where the gas radially behind the dense clump
is not photo-ionized or photo-heated by the source.  This produces
cometary and so-called elephant trunk structures that are also seen in
local star forming regions and have been discussed in detail since
\citet{Pottasch58}.  At a larger distance, the surrounding cosmic
structure is composed of intersecting or adjacent filaments and
satellite halos that breaks spherical symmetry.  The filaments and
nearby halos are optically thick and remain cool and thus the density
structures are largely unchanged.  The entropy of dense regions are
not increased by stellar radiation and will feel little negative
feedback from an entropy floor that only exists in the ionized IGM
\citep[cf.][]{Oh03}.  Ray-tracing allows for accurate tracking of
I-fronts in this inhomogeneous medium.  Radiation propagates through
the least optically thick path and generates champagne flows that have
been studied extensively in the context of present day star formation
\citep[e.g.][]{Franco90, Churchwell02, Shu02, Arthur06}.  In the
context of massive primordial stars, these champagne flows spread into
the voids and are impeded by the inflowing filaments.  The resulting
\ion{H}{2} regions have ``butterfly'' morphologies \citep{Abel99,
  Abel07, Alvarez06a, Mellema06, Yoshida07a}.  We also point out that
sources embedded in relic \ion{H}{2} largely maintain or increase the
ionization fraction.  Here the already low optical depth of the
recently ionized medium (within a recombination time) allows the
radiation to travel to greater distances than a halo embedded in a
completely neutral IGM.  The \ion{H}{2} regions become increasingly
anisotropic in higher mass halos.  We show an example of the
morphology of a \ion{H}{2} region near the end of the star's lifetime
in a dark matter halo with mass $1.4 \times 10^7 \Ms$ in Figure
\ref{fig:aniso}.

\subsection{Potential Caveats and Future Directions}

Although we have simulated the first generations of stars with
radiation hydrodynamic simulations, our methods have neglected some
potentially important processes and made an assumption about the Pop
III stellar masses.

One clear shortcoming of our simulations is the small volume and
limited statistics of the objects studied here.  However, it was our
intention to focus on the effects of Pop III star formation on
cosmological reionization and on the formation of an early dwarf
galaxy instead of global statistics.  The star formation only
simulations (SimA-RT and SimB-RT) converge to the similar averaged
quantities, e.g. ionized fraction, temperatures, star formation rates,
at the final redshift.  The evolution of these quantities differ
because of the limited number of stars that form in the simulations,
which then causes the evolution to depend on individual star formation
times.  This variance should be expected in the small volumes that we
simulate and should not diminish the significance of our results.

We have verified even in a 2.5-$\sigma$ peak that Pop III stars cannot
fully reionize the universe, which verified previous conclusions that
low-luminosity galaxies provide the majority of ionizing photons.
Furthermore, it is beneficial to study Pop III stellar feedback
because it regulates the nature of star formation in these galaxies
that form from pre-heated material.  Further radiation hydrodynamics
simulations of primordial star and galaxy formation with larger
volumes while still resolving the first star forming halos of mass
$\sim$$3 \times 10^5 \Ms$ will improve the statistics of early star
formation, especially in more typical overdensities, i.e. 1-$\sigma$
peaks, some of which could survive to become dwarf spheroidal galaxies
at $z = 0$.

In this work, we treated the LW radiation field as optically thin, but
in reality, \hh~produces a non-zero optical depth above column
densities of $10^{14} \> \mathrm{cm}^{-2}$ \citep{Draine96}.
Conversely, Doppler shifts of the LW lines arising from large velocity
anisotropies and gradient may render \hh~self-shielding unimportant up
to column densities of $10^{20} - 10^{21} \> \mathrm{cm}^{-2}$
\citep{Glover01}.  If self-shielding is important, it will lead to
increased star formation in low-mass halos even when a nearby source
is shining.  Moreover, \hh~production can also be catalyzed ahead of
I-fronts \citep{Ricotti01, Ahn07}.  In these halos, LW radiation will
be absorbed before it can dissociate the central \hh~core.  On the
same topic, we neglect any type of soft UV or LW background that is
created by sources that are cosmologically nearby ($\Delta z / z \sim
0.1$).  A soft UV background either creates positive or negative
feedback, depending on its strength \citep{Mesigner06}, and a LW
background increases the minimum halo mass of a star-forming halo
\citep{Machacek01, Yoshida03, OShea07, Wise07b}.  However in our
calculations, the lack of self-shielding, which suppresses star
formation in low-mass halos, and the neglect of a LW background, which
allows star formation in these halos, may partially cancel each other.
Hence one may expect no significant deviations in the SFRs and
reionization history if one treats these processes explicitly.

To address the incident radiation and the resulting UV background from
more rare density fluctuations outside of our simulation volume, it
will be useful to bridge the gap between the start of reionization on
Mpc scales to larger scale (10 -- 100 Mpc) simulations of
reionization, such as the work of \citet{Sokasian03}, \citet{Iliev06},
\citet{Zahn07}, and \citet{Kohler07}.  Radiation characteristics from
a volume that has similar overdensities as our Mpc-scale simulations
can be sampled from such larger volumes to create a radiation
background that inflicts the structures in our Mpc scale simulations.
Inversely, perhaps the small-scale evolution of the clumping factor,
filtering mass, and average temperature and ionization states can be
used to create an accurate subgrid model in large volume reionization
simulations.

Another potential caveat is the continued use of primordial gas
chemistry in metal enriched regions in the SN runs.  Our simulations
with SNe give excellent initial conditions to self-consistently
treating the transition to low-mass star formation.  In future work,
we plan to introduce metal-line and dust cooling models
\citep[e.g. from][] {Glover07, Smith07} to study this transition.

The one main assumption about Pop III stars in our calculations is the
fixed, user-defined stellar mass.  The initial mass function (IMF) of
these stars is largely unknown, therefore we did not want to introduce
an uncertainty by choosing a fiducial IMF.  It is possible to
calculate a rough estimate of the stellar mass by comparing the
accretion rates and Kelvin-Helmholtz time of the contracting molecular
cloud \citep{Abel02, OShea05}.  Protostellar models of primordial
stars have also shown that the zero-age main sequence (ZAMS) is
reached at 100 \Ms~for typical accretion histories after the star
halts its adiabatic contraction \citep{Omukai03, Yoshida06}.
Furthermore, we have neglected HD cooling, which may become important
in halos embedded in relic \ion{H}{2} regions and result in lower mass
($\sim$$30 \Ms$) metal-free stars \citep{OShea05, Greif06, Yoshida07b}.
Based on accretion histories of star forming halos, one can estimate
the ZAMS stellar mass for each halo and create a more self-consistent
and ab initio treatment of Pop III star formation and feedback.

\section{SUMMARY}
\label{sec:summary}

We conducted three radiation hydrodynamical, adaptive mesh refinement
simulations that supplement our previous cosmological simulations that
focused on the hydrodynamics and cooling during early galaxy
formation.  These new simulations concentrated on the formation and
feedback of massive, metal-free stars.  We used adaptive ray tracing
to accurately track the resulting \ion{H}{2} regions and followed the
evolution of the photo-ionized and photo-heated IGM.  We also explored
on the details of early star formation in these simulations.  Theories
of early galaxy formation and reionization and large scale
reionization simulations can benefit from the useful quantities and
characteristics of the high redshift universe, such as SFR and IGM
temperatures and ionization states, calculated in our simulations.
The key results from this work are listed below.

\medskip

1. SFRs increase from $5 \times 10^{-4}$ at redshift 30 to $6 \times
10^{-3}$ \sfr~at redshift 20 in our simulations.  Afterwards the SFR
begins to have a bursting nature in halos more massive than $10^7 \Ms$
and fluctuates around $10^{-2}$ \sfr.  These rates are larger than the
ones calculated in \citet{Hernquist03} because our simulation volume
samples a highly biased region that contains a 2.5-$\sigma$ density
fluctuation.  The associated emissivity from these stars increase from
1 to $\sim$100 ionizing photons per baryon per Hubble time between
redshifts 15 and 30.

2. In order to provide a comparison to semi-analytic models, we
calculate the star formation efficiency to be $\sim$$6 \times 10^{-4}$
averaged over all redshifts and the simulation volume.  For Pop III
star formation, this is a factor of two lower than stars that are not
affected by feedback \citep{Abel02, Bromm02, Yoshida06, OShea07}.

3. Shock waves created by D-type I-fronts expel most of the gas in the
host halos below $\sim$$5 \times 10^6 \Ms$.  Above this mass,
significant outflows that are still bound to the halo are generated.
This feedback creates a dynamical picture of early structure
formation, where star formation is suppressed in halos because of this
baryon depletion, which is more effective than UV heating or the
radiative dissociation of \hh.

4. We see three instances of induced star formation in halos with
masses $\sim 3 \times 10^6 \Ms$.  Here a star forms as a SN blast wave
overtakes an overdensity created by an ionization front instability.
\hh~formation is catalyzed by additional free electrons in the relic
\ion{H}{2} region and in the SN blast wave \citep{Ferrara98}.

5. As star formation occurs regularly in the simulation after redshift
25, four (six) ionizing photons are needed per sustained hydrogen
ionization.  As the most massive halo becomes larger than $\sim$$10^7
\Ms$ in the simulations without SNe, \ion{H}{2} regions become trapped
and ionizing radiation only escapes into the IGM in small solid
angles.  Hence the number of photons per effective ionization
increases to 15 (50).  In SimB-SN, stellar radiation from induced star
formation have an escape fraction of nearly unity, which occur four
times in the calculation.  This allows the IGM to remain ionized at a
volume fraction 3 times higher than without SNe.  Similarly, the
ionizing photon to ionization ratio also stays elevated at 10:1
instead of decreasing in the calculations with star formation only.

6. Our simulations that include star formation and \hh~formation
capture the entire evolution of the clumping factor that is used in
semi-analytic models to calculate the effective enhancement of
recombinations in the IGM.  We showed that clumping factors in the
ionized medium fluctuate around the 75\% of the values found in
adiabatic simulations.  They evolve from unity at high redshifts and
steadily increase to $\sim$4 and 3.5 with and without SNe at $z = 17$,
respectively.  Photo-evaporation from stellar feedback causes the
decrease of the clumping factor.

7. We calculated the Jeans filtering mass with the volume-averaged
temperature only in fully and partially ionized regions, which yields
a better estimate than the temperature averaged over both ionized and
neutral regions.  The filtering mass depends on the thermal history of
the IGM, which mainly cools through Compton cooling.  It increases by
two orders of magnitude to $\sim$$3 \times 10^7 \Ms$ at $z \sim 15$.
It describes the minimum mass a halo requires to collapse after
hosting a Pop III star.  For halos forming their first star, the
minimum halo mass is regulated by the LW background \citep{Machacek01}
and photo-evaporation \citep[e.g.][]{Haiman01}.

\medskip

Pop III stellar feedback plays a key role in early star formation and
the beginning of cosmological reionization.  The shallow potential
wells of their host halos only amplify their radiative feedback.  Our
understanding of the formation of the oldest galaxies and the
characteristics of isolated dwarf galaxies may benefit from including
the earliest stars and their feedback in galaxy formation models.
Although these massive stars only partially reionized the universe,
their feedback on the IGM and galaxies is crucial to include since it
affects the characteristics of low-mass galaxies that are thought to
be primarily responsible for cosmological reionization.  Harnessing
observational clues about reionization, observations of local dwarf
spheroidal galaxies, and numerical simulations that accurately handle
star formation and feedback may provide great insight on the formation
of the first galaxies, their properties, and how they completed
cosmological reionization.

\acknowledgments

The quality and robustness of this paper was improved through the
feedback of an anonymous referee.  This work was supported by NSF
CAREER award AST-0239709 from the National Science Foundation and
partially supported in part by the National Science Foundation under
Grant No. PHY05-51164.  J.~H.~W. thanks Renyue Cen for helpful
discussions.  We benefited from fruitful conversations with Marcelo
Alvarez.  We are grateful for the continuous support from the
computational team at SLAC.  We benefited the hospitality of KITP at
UC Santa Barbara, where this work was completed.  We performed these
calculations on 16 processors of a SGI Altix 3700 Bx2 at KIPAC at
Stanford University.

{}

\end{document}